# The impact of interface modification on the behavior of phenyl alcohols within silica templates

Natalia Soszka[a], Magdalena Tarnacka[b*], Barbara Hachuła[a], Monika Geppert-Rybczyńska[a], Krystian Prusik[c], Kamil Kamiński[b]

[a] *Institute of Chemistry, University of Silesia in Katowice, Szkolna 9, 40-006 Katowice, Poland*

[b] *August Chełkowski Institute of Physics, University of Silesia in Katowice, 75 Pułku Piechoty 1a, 41- 500 Chorzów, Poland*

[c] *Institute of Materials Engineering, University of Silesia in Katowice, 75 Pułku Piechoty 1a, 41-500 Chorzów, Poland*

*Corresponding author: magdalena.tarnacka@us.edu.pl

**Abstract**

Herein, thermal, dynamical properties, host-guest intermolecular interactions, and wettability of a series of monohydroxy phenyl-substituted alcohols (PhAs) infiltrated into native and silanized silica mesopores (d = 4 nm) were investigated by means of Dielectric and Infrared (IR) Spectroscopy, Differential Scanning Calorimetry as well as the contact angle measurements. Calorimetric data showed the occurrence of the two glass transition temperatures, $T_g$. Importantly, around the one detected at higher temperatures ($T_{g,interfacial}$), strong deviation in the temperature evolution of the relaxation time of the main process was observed for all systems. Moreover, an additional process unrelated to the mobility of interface layer and core molecules was revealed most likely connected to either SAP or 'new' confinement-induced nanoassociates. Further, IR investigations showed that the applied nanoconfinement had little impact on hydrogen bonds' strength, but it influenced the HBs distribution (including 'new' population of HB) and the degree of association. Additionally, for the first time, we calculated the activation energy values of the dissociation process for PhAs in mesopores, which turned out to be lower with respect to those estimated for bulk samples. Thus, our research clearly showed the impact of the spatial geometrical restriction on the association process in alcohols having significant steric hindrance.

## I. Introduction

Geometric nanoconfinement offers a versatile platform for modifying the structural and dynamical properties of soft matter systems, making them highly relevant for nanoscale applications. Previous research has revealed that soft matter adjacent to a solid and near-surface can exhibit interfacial regions of reduced density [1–3], interfacial freezing [4–6], interfacial melting [7–10], molecular layering [11–15], molecular orientation [16], or specific lateral molecular arrangements [17]. Consequently, significant deviations in their properties, such as flow and intermolecular entanglements, local chain mobility, glass transition temperature, diffusion, tautomerization, or crystallization, is noted when compared to bulk samples. [18–27] Due to these facts, confined systems hold promise for diverse industrial applications, including nanolithography, lubrication, paints, surface treatments, and elastic membranes. [28]

Among many types of nanorestrictions, silica mesoporous materials, such as MCM-41 or SBA-15, possess the O-H groups (silanol groups ($\equiv$Si−OH)) inside pores that can easily interact through H-bonds with lattice oxygen. To successfully tailor these materials for nanotechnology applications, the membrane characteristic is often adjusted by altering the pore diameter and/or surface properties. The latter includes post-synthesis modifications, typically involving silanizing agents like as methoxytrimethylsilane, trimethyl chloride [29], mercaptopropyltrimethoxysilane [30], hexamethyldisilazane [31], or chlorotrimethylsilane [32], which change the surface of the silica from hydrophilic to hydrophobic by replacing the O-H groups with organic groups via O-Si-C covalent bonds. This alteration affects pore structure, which determines the nature of host-guest interactions. To illustrate that, one can refer to the studies on water in hydrophilic and hydrophobic nanochannels. In the former case, IR spectroscopy revealed a predominant ice-like structure because of comparably strong interactions between the water molecules and the pore wall (interface interactions) [29], while for the latter one, water behaved more liquid-like (bulk-like), as water molecules interacted less with the silica surface (fewer interaction sites at the pore wall) and more with each other [29]. Similar trends were also observed for aliphatic mono- and polyhydroxy alcohols [31,33] infiltrated into nanopores, where hydrophobization of the silica surface hinders H-bonds between the pore wall and the guest molecules, making the dynamics of confined liquid more bulk-like. [31] The effects related to the reduction of surface interactions were also seen for hydroxyl-terminated polymers infiltrated into modulated anodic alumina (AAO) templates. [34] In this case, the factor responsible for the disturbance of interaction at the polymer/membrane interface was the roughness of applied nanotemplates. Importantly, similar results were also reported for PhAs infiltrated within nanoporous anodic aluminum oxide (AAO) membranes having constant and varying pore diameter ($d$ = 10-40 nm). Samples within ordinary AAO templates revealed a pronounced confinement effect on $T_g$ and structural relaxation process, whereas alcohols in the latter templates showed a bulk-like behavior in the whole range of studied temperatures. [35] Herein, one can also mention the papers where authors explored the impact of the position of the hydroxyl group (primary, secondary, tertiary) in the molecular skeleton of monohydroxy alcohol (MA) being nanospatially restricted on their supramolecular organization and molecular dynamics. [36–39]

In this paper, we investigated the effect of surface interactions on the behavior of a series of phenyl-substituted monohydroxy alcohols (PhAs) from ethanol to pentanol (see their chemical structure in **Scheme 1**) incorporated into mesoporous (un)treated silica templates characterized by a pore diameter, $d$ = 4 nm, by means of Broadband Dielectric Spectroscopy (BDS), Differential Scanning Calorimetry (DSC), and Fourier Transform Infrared Spectroscopy (FTIR).

## II. Materials and methods

**Materials.** 2-phenyl-1-ethanol (2Ph1E), 3-phenyl-1-propanol (3Ph1P), 4-phenyl-1-butanol (4Ph1B), and 5-phenyl-1-pentanol (5Ph1P) with purity higher than 98% were supplied by Sigma Aldrich. Their chemical structures are shown in **Scheme 1**. The native and silanized silica mesopores of pore diameter, $d$=4 nm (see **Scheme 2**), were produced by electrochemical etching of silicon wafers and subsequent thermal oxidation. Details about the fabrication of silica templates and their further modification are presented in Ref. [40,41].

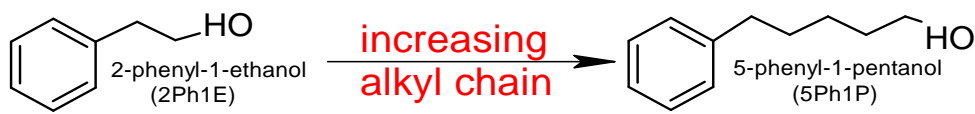

**Scheme 1.** The chemical structures of investigated phenyl alcohols.

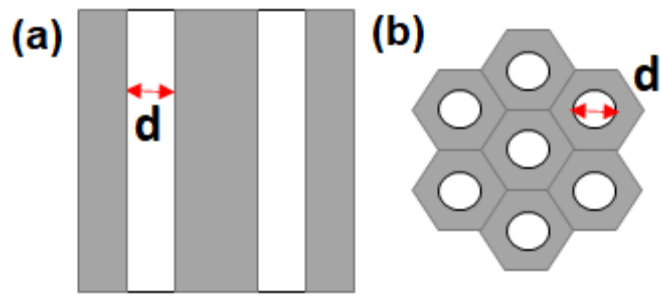

**Scheme 2.** The schematic structure of (a) cross-section and (b) the top of applied silica mesopore templates ($d$ denotes the pore diameter).

**Sample preparations.** Prior to filling, the silica membrane was dried in the oven at T = 373 K under vacuum in order to remove any volatile impurities from the nanochannels, and alcohols were subjected to freezing in liquid nitrogen to remove water. After cooling, membranes were infiltrated with alcohols. Then, the alcohol-membrane systems were maintained in a vacuum ($10^{-2}$ bar) to allow the liquid to flow inside via capillary forces. After the infiltration process, the excess of sample on the surface was removed. The pore diameter of silica nanopores was confirmed by SEM measurements (see **Figure S1** in **Supporting Material, SM file**). The filling degree reaches ∼90%. Those values are calculated considering the porosity of membranes and the assumption that the density of the infiltrated material does not change along the pore radius and that the shape of the pore is cylindrical.

**Differential Scanning Calorimetry (DSC).** Calorimetric measurements were carried out by a Mettler-Toledo DSC apparatus equipped with a liquid nitrogen cooling accessory and an HSS8 ceramic sensor (heat flux sensor with 120 thermocouples). Temperature and enthalpy calibrations were performed by using indium and zinc standards. The sample was prepared in an open aluminum crucible (40 μL) outside the DSC apparatus. Samples were scanned at various temperatures at a constant heating rate of 10 K/min.

**Surface Tension and Contact Angle Measurements.** The surface tension of liquids, $\gamma_L$ (the pendant drop method), and contact angle, $\theta$, was measured by the Drop Shape Analysis 100S Krüss Tensiometer, GmbH Germany, with the Advance software. The contact angle measurements were performed at nine temperatures in the wide temperature range, $T = 263 - 298$ K. Thanks to the measuring chamber adopted to work in the wide temperature range with humidity control, it was possible to measure contact angle below the freezing point of water. The general description of procedures has been presented previously. [42] The contact angle measurements on the solid smooth surfaces have been repeated a dozen or more times. The temperature measurement uncertainty was ± 0.1 K. The precision of contact angle measurements was 0.01°, and the estimated uncertainty was not more than ± 2°, whereas the uncertainty of the surface tension was ± 0.1 mN/m. The surface energy, $\gamma_S$, for native silica was $\gamma_S$ = 67.6 mJ/m² with dominant non-dispersive part equal 66.6 mJ/m². For silanized respective value was $\gamma_S$ = 25.3 mJ/m² with dominant-dispersive part 22.8 mJ/m². [43]

**Broadband dielectric spectroscopy (BDS).** Measurements were carried out on heating after a fast quenching of the liquid state in a wide range of temperatures (175 – 243 K) and frequencies ($10^{-1} - 10^6$ Hz) using a Novocontrol spectrometer equipped with Alpha Impedance Analyzer with an active sample cell and Quatro Cryosystem. Dielectric measurements of bulk samples were performed in a parallel-plate cell (diameter: 15 mm, gap: 0.1 mm) as described in Ref. [44]. Silica membranes filled with studied alcohols were also placed in a similar capacitor (diameter: 10 mm, membrane thickness: 0.05 mm). [40,41] Nevertheless, the confined samples are a heterogeneous dielectric consisting of a matrix and an investigated compound. Because the applied electric field is parallel to the long pore axes, the equivalent circuit consists of two capacitors in parallel composed of $\varepsilon^*_{compound}$ and $\varepsilon^*_{templates}$. Thus, the measured total impedance is related to the individual values through $1/Z^*_c = 1/Z^*_{compound} + 1/Z^*_{templates}$, where the contribution of the matrix is marginal. The measured dielectric spectra were corrected according to the method presented in Ref. [45].

The normalized α-loss peak for bulk was fitted to the one-sided Fourier transform of the Kohlrausch-Williams-Watts (KWW) function (dotted line in **Figures 2(c,f)**):

$$\Phi_{KWW}(t) = \exp[-(\tau/\tau_\alpha)^{\beta_{KWW}}] \quad (1)$$

We found that for bulk PhAs, the fractional exponent, $\beta_{KWW}$, which describes the α-relaxation time distribution's nonexponentiality, is equal to $\beta_{KWW}$ = 0.9.

In order to obtain relaxation times, $\tau$, and also shape parameters of relaxation peaks, the dielectric data were fitted using the Havrilak-Negami (HN) function [46]:

$$\varepsilon(\bar\omega)'' = \frac{\sigma_{DC}}{\varepsilon_0 \bar\omega} + \frac{\Delta\varepsilon}{[1+(i\bar\omega\tau_{HN})^{\alpha_{HN}}]^{\beta_{HN}}} \quad (2)$$

where $\sigma_{DC}$ is the dc-conductivity term, $\varepsilon_0$ is the vacuum permittivity, $\Delta\varepsilon$ is the dielectric strength, $\bar\omega$ is the angular frequency, $\tau_{HN}$ describes HN relaxation time and $\alpha_{HN}, \beta_{HN}$ are the shape parameters.

The temperature dependence curves of the relaxation times of bulk Debye-like process were fitted using a Vogel-Fulcher-Tamman (VFT) function: [47–49]

$$\tau_D = \tau_\infty \exp\left(\frac{D_T T_0}{T - T_0}\right) \qquad (3)$$

where, $\tau_\infty$ is the relaxation time at finite temperature, $T_0$ is the temperature when $\tau$ goes to infinity, and $D_T$ is the fragility parameter. $T_g s$ of the bulks were determined as the temperature at which $\tau = 100$ s by extrapolating the VFT fits. On the other hand, to determine $T_{g,core}$ for the confined systems, their $\tau_{Dom}(T)$ dependences below $T_{g,interfacial}$ (after the kink), were fitted with either VFT or the Arrhenius equation (dependently on the sample):

$$\tau_D = \tau_\infty \exp\left(\frac{\Delta E}{k_b T}\right) \qquad (4)$$

where, $k_b$ is the Boltzmann constant, and $\Delta E$ is the activation energy.

**Fourier Transformed Infrared Spectroscopy (FTIR).** FTIR spectra of bulk and confined samples were recorded on Nicolet iS50 FTIR spectrometer (Thermo Scientific). The spectra were collected in the spectral region 4000 – 1300 cm$^{-1}$ by averaging 16 scans with a spectral resolution of 2 cm-1. Due to the absorption of CaF$_2$ windows and saturation of silica membranes, the region below 1300 cm$^{-1}$ was not analyzed. The confined sample's spectra were measured with an 'empty' silica membrane spectrum as a background in order to eliminate any influence of the template on the spectra of incorporated alcohol (**Figure S2**). The measurements were taken in a broad range of temperatures (T = 297 – 153 K) every 4 K with a temperature rate of 4 K/min. To record the temperature-dependent spectra, the Linkam THMS 600 stage was used. Throughout the whole procedure, the spectrometer was purged with liquid nitrogen.

## III. Results and Discussion

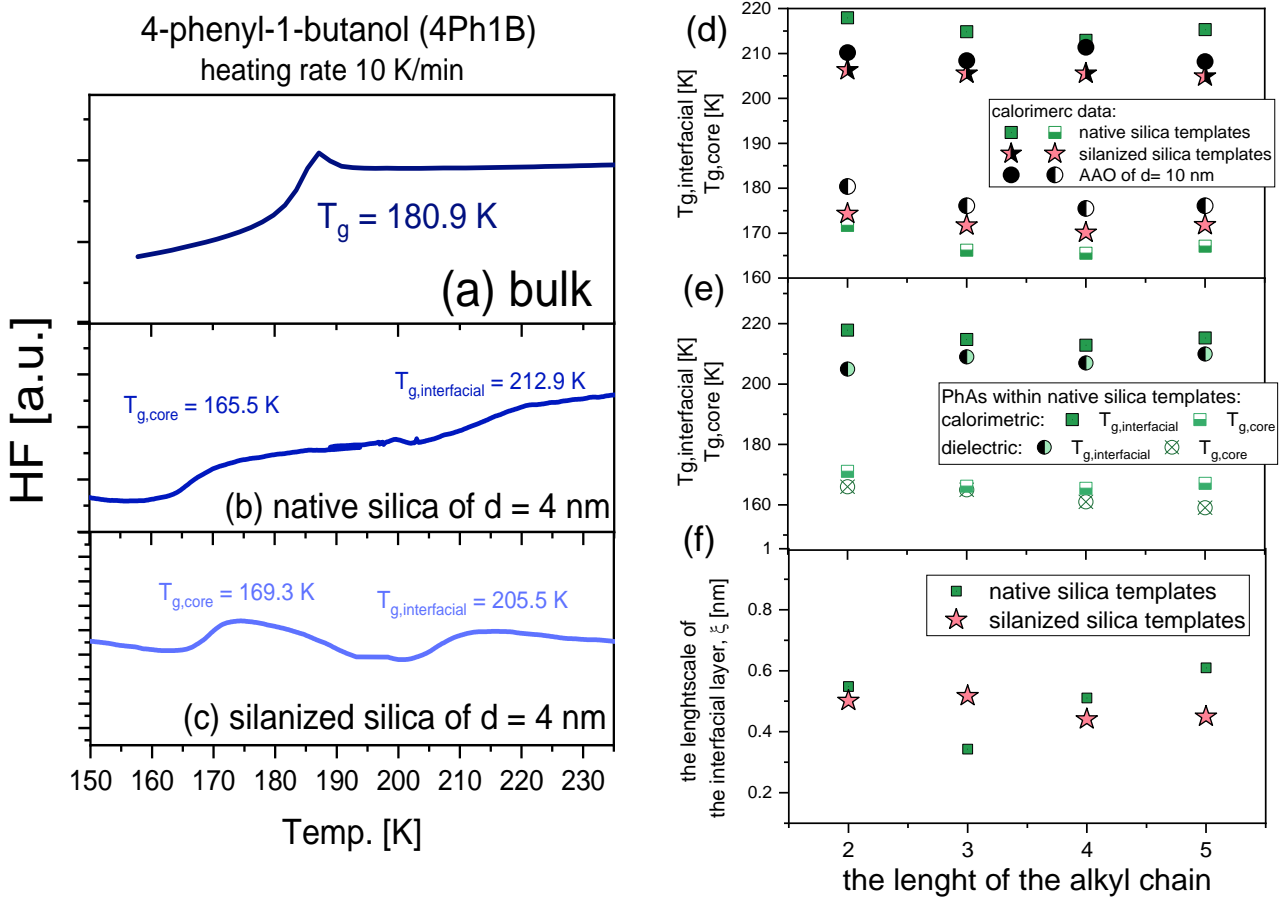

**Figure 1**. (a-c) DSC thermograms obtained for bulk and confined samples of 4Ph1B, (d) The glass transition temperatures (determined from calorimetric measurements) of all confined samples vs. the length of the alkyl chain of the examined PhA; (e) Comparison of the glass transition temperatures obtained from BDS and DSC measurements, for all studied infiltrated samples. The uncertainty of $T_g$ determination is ± 2 K for all presented values. (f) The thickness of the interfacial layer ($\xi$) plotted as a function of the length of the alkyl chain.

Representative DSC curves recorded for 4Ph1B, bulk, and samples infiltrated into various silica membranes are presented in **Figures 1(a-c)**. All bulk samples exhibit the presence of one glass transition (manifested by a characteristic endothermic jump of heat capacity). Moreover, $T_g$s reveal a slight 'odd-even' effect with the change in molecular weight.[44] Values of bulk glass transition temperatures are listed in **Table S1.** Instead, for PhAs incorporated within various silica membranes characterized by $d = 4$ nm, two (double) glass transitions (DGT, manifested as two endothermic signals) located above and below the bulk $T_g$. One can stress that herein, the DGT scenario was recorded for all examined confined samples, regardless of the length of the alkyl chain of PhAs and the type of templates (native/silanized). It should be noted that the double glass transition was previously reported for a variety of glass formers under confinement generated by the application of porous templates, including associating materials, [50–55], and is currently considered as a common feature of this type of material. The presence of DGT is assumed to originate from the confinement-induced distribution of interactions within the systems due to the presence of (interacting) interface leading to the heterogeneity in terms of molecular dynamics as well as packing density. Those differences can be described by the simple 'two-layer' (or 'core-shell') model [52,56], which implies that you can distinguish previously mentioned 'interfacial' molecules of reduced mobility with the glass transition occurring at higher temperatures ($T_{g,interfacial}$), and the ones located at some distance from the interface (more in the middle of the nanochannels) labeled as the 'core' fraction characterized by lower values of the glass transition temperatures ($T_{g,core}$). One can recall previous data for PhAs infiltrated within various AAO templates of different $d$, which also revealed the presence of DGT.[35] Nevertheless, it should be highlighted that, surprisingly, herein, we also observed the DGT for MAs incorporated within modified (silanized) mesoporous. One can recall that up to now, it was assumed that the lack of specific interactions between confined molecules and the interface would result in the absence of a defined interfacial layer. Although recent studies of polar van der Waals liquid, S-methoxy-PC, [57] incorporated within AAO membranes of different pore sizes and modified surface clearly shows that it is not the case, and even for silanized hydrophobic surface, the trace of DGT (especially the presence of $T_{g,interfacial}$) can be seen. Values of determined $T_g$s are presented in **Figure 1(d)** and **Table S1**. As shown, $T_{g,interfacial}$ and $T_{g,core}$ decrease slightly ($\Delta T_g \sim 5K$) with the elongation of the alkyl chain for the samples infiltrated into both types of applied templates. Nevertheless, there is a pronounced difference in $T_g$s between materials within either native or silanized silica templates, where the difference between both $T_g$s (defined as $\Delta T_g = T_{g,interfcial} - T_{g,core}$) is higher for alcohols within native silica templates in comparison to those within the treated (modified) ones. This possibly originates from the specific H bonding interactions between native silica and alcohol, while in the silanized pores, these specific interactions are strongly reduced.

Next, in order to quantify the volume of the interfacial layer dependently on the surface interactions, we calculated the length scale of the interfacial layer, $\xi$, as follows [54]:

$$\xi = \frac{d}{2}\left[1 - \left(1 - \frac{\Delta C_{p,interfacial}}{\Delta C_{p,interfacial}+\Delta C_{p,core}}\right)^{1/2}\right] \quad (5)$$

where $\Delta C_{p,i}$s are heat capacity changes at corresponding $T_{g,i}$. Note that Eq. (5) was applied under the following conditions: the volume of the material in the surface layer is proportional to the step change of its heat capacity; the density of the confined sample is constant along the pore radius; and the pore is cylindrical. Additionally, it should be highlighted that Eq. (5) is a simple mathematical model, assuming direct proportionality between heat capacity and number of molecules, and it does not take into account any variation in the density, roughness, or curvatures of applied templates. This might lead to some overestimation of $\xi$, which we are aware of. The determined values of $\xi$ for all examined samples plotted as a function of the alkyl chain are shown in **Figure 1(f)**. As illustrated, the interfacial layer reaches the length scale of approx. $\xi \sim 0.5$ nm for all samples incorporated in native or silanized porous template. For comparison, one can mention that for PhAs infiltrated within AAO templates of significantly higher pore size, $d$ = 10-40 nm, $\xi \sim 2$-7 nm.[35] Such a high value of $\xi$ seems to indicate up to 10 molecular layers, which might be achieved by the formation of some associating structure (most likely loosely packed) near/at the interface. Moreover, one can add that for other MAs, i.e., 2-ethyl-1-hexanol, 2-ethyl-1-butanol, and 5-methyl-3-heptanol, infiltrated into native silica templates of comparable pore size ($d$ = 4-8 nm), the length scale of interfacial layer reaches $\xi \approx 0.5$ -1.4 nm. [37]

**Table 1**. Contact angle, $\theta$, at 298 K and at 258 K, as well as the surface tension, $\gamma_L$.

| Sample | $\gamma_L$ at 298 K [mN/m][a] | Native silica | | Silanized silica | |
|---|---|---|---|---|---|
| | | $\theta$ at 298 K [°] | $\theta$ at 258 K [°] | $\theta$ at 298 K [°] | $\theta$ at 258 K [°] |
| 2Ph1E | 38.84 | 30.2 | 22.2 | 44.0 | 28.3 |
| 3Ph1P | 38.28 | 28.4 | 21.8 | 42.0 | 28.6 |
| 4Ph1B | 37.26 | 28.2 | 22.0 | 39,2 | 26.6 |
| 5Ph1P | 37.01 | 28.9 | 22.2 | 39.8 | 26.9 |

[a] Values of $\gamma_S$ were taken from Ref. [44].

In order to understand why we observe so clearly DGT in recorded thermograms and rather comparable length scale of the interfacial layer in alcohols infiltrated within native and modified silica, we carried out additional measurements of the contact angle, $\theta$, [43,58,59] which can be used to quantify wettability. Note that the smaller $\theta$, indicates the better spreading out of a liquid drop on the examined surface (and thus, better wettability). Values of determined $\theta$ for all examined PhAs are shown in **Table 1** and **Figure S3**. As can be seen, all materials are characterized by similar contact angles independently of the examined surface; however, it can be observed that in the case of silanized one, values of $\theta$ are higher, as can be expected due to differences in the surface's interactions. Nevertheless, contrary to the expectations, both surfaces can be considered as hydrophilic since, for all cases $\theta$ <90º. Therefore, one can state that although there is some difference in the wettability of the material in contact with the native and modified silica, it is not significant. Moreover, it should be pointed out that the contact angle decreases with lowering temperature, reaching comparable values $\theta \sim 22º$ and $\theta \sim 28º$ for PhAs spread on the respectively native and silanized silica surfaces at $T = 266$ K (see **Table 1** and **Figure S3)**. That means that at low temperatures, wettability and intermolecular interactions between materials and membrane pore walls are comparable for all systems independently of the applied type of porous templates. Therefore, one might assume that specific H-bonding interactions with pore walls do not play a significant role in the enhancement of the wettability of the PhA in native silica with respect to the silanized one. Nevertheless, a noticeable shift in the $T_{g,interfacial}$ to the higher temperature for the samples inlitrated into former membranes with respect to the modified templates is most likely due to the formation of the strong H bonds between host and guest molecules which are not present in the latter system.

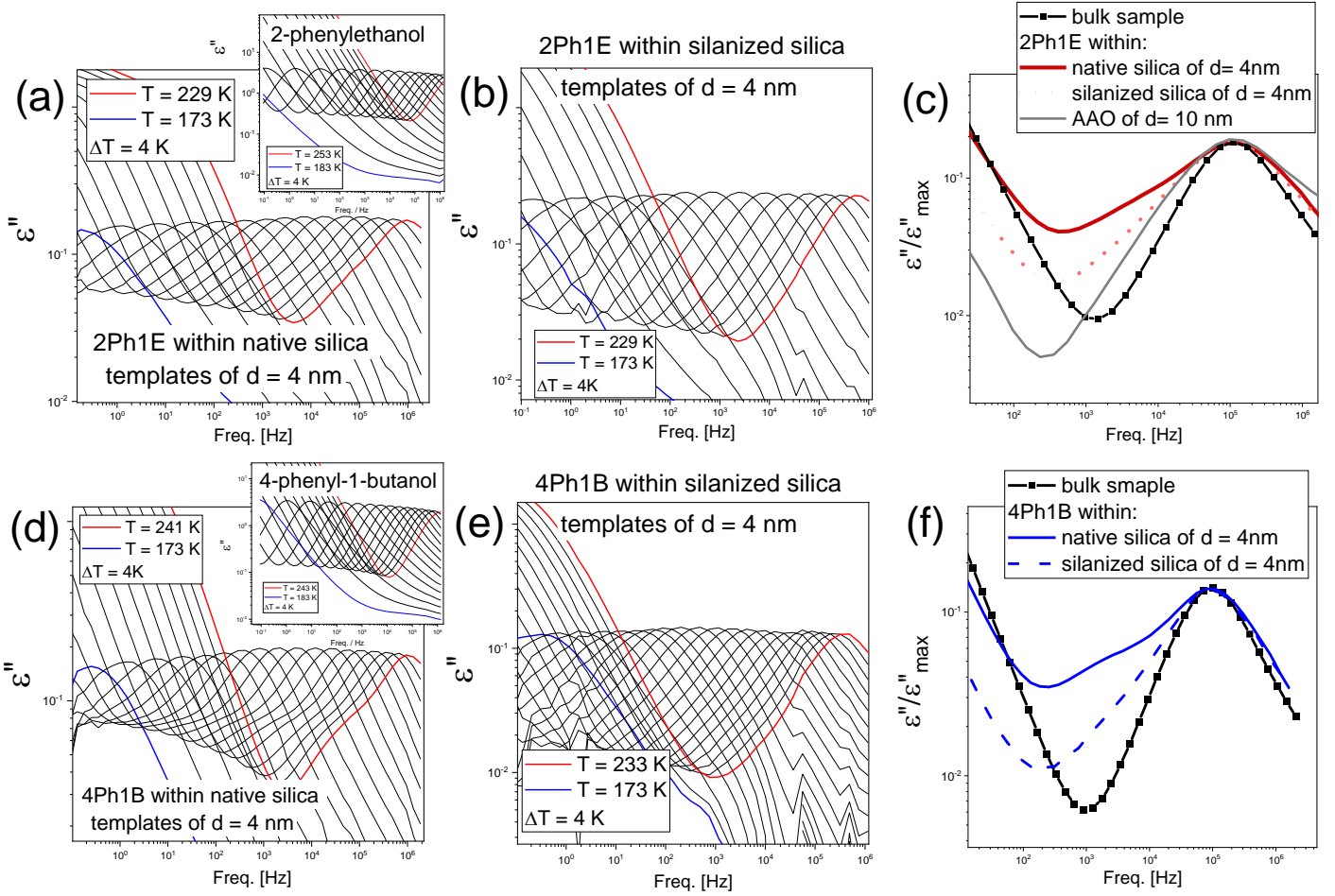

**Figure 2**. (a,b,d,e) Dielectric loss spectra of 2Ph1E (a,b) and 4Ph1B (d,e) infiltrated within native (a,d) and silanized (b,e) silica templates. As the insets in panels (a,d), dielectric data for the bulk systems are shown. (c,f) Comparison of dielectric loss peaks obtained for all measured samples at $f = 10^5$ Hz.

Furthermore, we performed comprehensive dielectric measurements. **Figure 2** illustrates representative dielectric loss, $\varepsilon''$ spectra obtained for 2-phenyl-1-ethanol (2Ph1E) and 4-phenyl-1-butanol (4Ph1B) incorporated within native and silanized silica templates of $d = 4$ nm. Data for the non-confined compound are presented as insets in **Figures 2(a,d)**. As shown, spectra obtained for bulk samples are composed of two relaxation processes, dc-conductivity at lower frequencies (related to the ion transport) and a prominent relaxation process at higher frequencies that dominates the recorded spectra (characterized by a Kohlrausch–Williams–Watts (KWW) stretched exponent, $\beta_{KWW}$ ~ 0.90, suggesting the Debye-like response, see **Figures 2(c,f)**). [44] Therefore, in this paper, we would label this process as Debye-like ($D$) one for the macroscale systems. Note that the loss spectra of various MA often exhibit the presence of a prominent Debye-like peak (originating from the formation of associating structures [60]), whereas the structural, $\alpha$, process (cooperative motions of molecules) often manifests itself as an excess wing in the high-frequency region. [37,41,61,62] Herein, we observed only one prominent symmetric mode with no sign of any additional process. Therefore, at first, this single dominant relaxation observed in the dielectric response of PhAs was interpreted as a *genuine* $\alpha$-mode. [63] However, recent studies on phenyl-substituted alcohols by a combination of BDS with different experimental techniques (i.e., either photon correlation spectroscopy, PCS, or mechanical measurements [62,64–66]) clearly show that this relaxation observed in the case of PhAs is, in fact, the superposition of the two processes (a slow Debye-like and $\alpha$-one, resulting from both cross correlation between dipole-dipole and self-dipole correlation, respectively) but observed as only one relaxation peak due to their similar time scale. [62]

Interestingly, in the case of PhAs infiltrated in various silica membranes, one can observe a different scenario. Firstly, for alcohols within native (untreated) templates, the presence of three relaxation processes can be detected (instead of two modes observed in bulk). An additional relaxation process emerges in the middle-frequency range on the low-frequency side of the main relaxation peak. Importantly, it becomes more prominent with an increase in the alkyl chain of examined compounds, see **Figures 2(c,f)**. On the other hand, it is not resolved for alcohols within silanized (treated) templates. Taking those observations into consideration, one can assume that the additional process might be, in fact, related to the motions of the molecules in close proximity to the interface, which strongly interacts with the pore walls (so-called interfacial molecules). [37,67] Therefore, often, this process is denoted as the interfacial one. Note that it is assumed that this mobility is no longer present within materials confined within treated silica templates due to a lack of specific interactions between PhA molecules and interface as a result of the silanization. However, taking into account the above-mentioned calorimetric data, one can also expect to observe the interfacial process also in the case of silanized systems. In this context, one can briefly remind that this process was not previously reported for alcohols incorporated in AAO templates of various pore sizes [35], which was assigned to the time scale of the mass exchange between interfacial and core molecules and experiment time. [68] It is assumed that when the exchange between both fractions is faster than the time of the experiments, the interfacial process can be detected (and is absent if not). Thus, one can assume that the lack of a well-resolved additional process of PhAs within silanized mesoporous might be related to the differences between the time scale of the mass exchange between interfacial and core molecules and experiment time when compared to the MAs within native templates. [68,69]

**Figures 2(c,f),** representing the comparison of the shape of the main relaxation process for all examined systems, is shown in comparison to the bulk. The presented spectra were shifted to superposed at the same relaxation time as the main relaxation process. As illustrated, the dominant relaxation peak observed for all examined confined systems is broader (especially in the low-frequency region) than those of the bulk sample. This agrees with the previous data reported for PhAs infiltrated within various AAO templates (data for alcohols within AAO membranes of $d = 10$ nm were added). One can mention that this behavior is a common feature of porous materials, observed for both van der Waals and associating liquids, [37,70–73] regardless of the material (AAO/silica) applied porous membranes are made of. This broadening of the distribution of the relaxation times is often discussed as a result of an increase in the heterogeneity of the molecular mobility observed in the confined systems, most likely induced by the presence of the interface (in terms of the introduction of additional interactions within the examined systems). [67,74] One can add that for associating materials, also changes in the hydrogen bond population under confinement might be taken into consideration. [62,75] At this point, two issues should be highlighted. First, the observed loss peak under confinement is no longer a Debye-like relaxation; therefore, in the case of confined samples, this mode would be referred to as a dominant/main relaxation process. Second, the significant broadening of the low-frequency side of the main relaxation peak for PhAs within native silica templates suggests emerging a new mode, which might be assigned to the interfacial process as discussed above. However, when examining the spectra collected for PhAs within silanized (treated) templates, one can also see that the interfacial process can be observed for those systems, but it has a significantly lower amplitude when compared to the materials within native membranes. [40,41] This might imply that even if the specific interactions at the interface are suppressed due to performed surface modification, some van der Waals interactions near the surface can still be present, contributing to the formation of the molecule fraction characterized by reduced mobility in the proximity of the more hydrophobic surface. Note that the interfacial process is also affected by the magnitude of dipole moment libration of immobilized molecules as well as the long-distance correlation between dipoles. Nevertheless, it should be mentioned that up to now, no presence of the interfacial process was reported for materials infiltrated within modified silica templates. [40,41,76,77] Therefore, the question arises about the origin of the observed additional relaxation process.

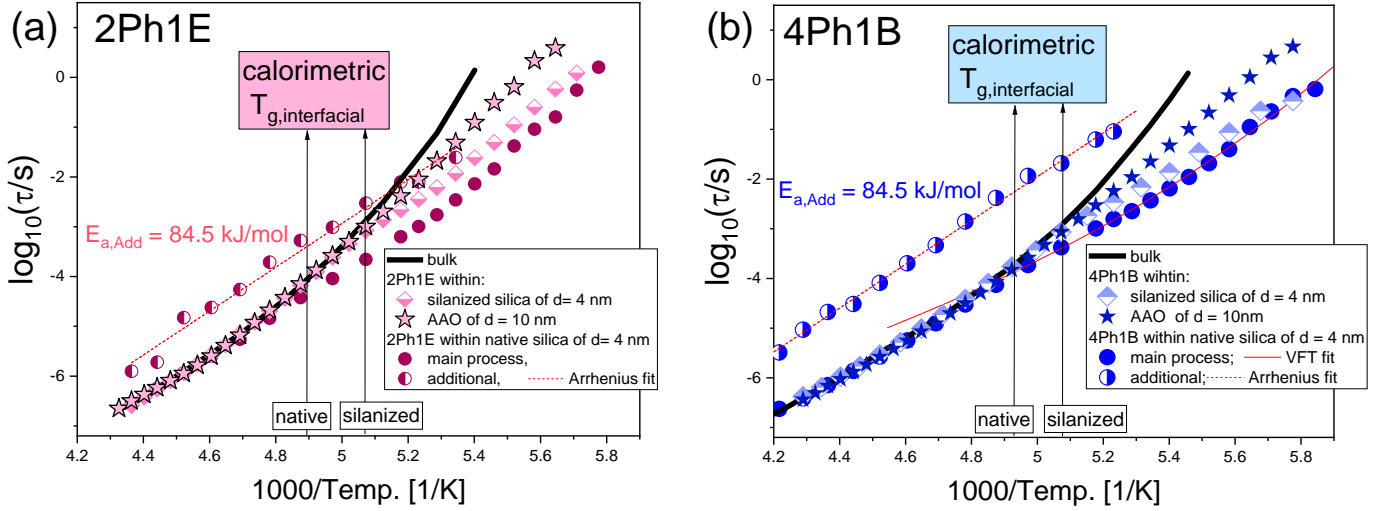

**Figure 3**. Temperature dependences of relaxation time ($\tau$) obtained for 2Ph1E (a) and 4Ph1B (b), both bulk and confined. For comparison, data for PhAs within AAO templates characterized by $d$ = 10 nm are shown (taken from Ref. [35]).

To provide additional information, in the next step, we determined the temperature dependence of the relaxation times, $\tau$, of observed relaxation peaks, the dominant ($\tau_{Dom}$) and additional ($\tau_{Add}$) ones located at high- and middle-frequency regions, respectively. For that purpose, the dielectric data were fitted using (one or two) Havrilak-Negami (HN) function(s) with an additional dc-conductivity term (see **Materials and Methods**). [78] Determined in this way relaxation times were plotted versus inverse temperature and shown in **Figure 3**. As illustrated, there are no differences between the $\tau_{Dom}(T)$-dependences of examined systems (either bulk or infiltrated) at the 'high-temperature' region, while further cooling revealed a systematic change in the slope of $\tau_{Dom}(T)$ of infiltrated materials from VFT-like of bulk characteristic to the more Arrhenius-like at some specific temperature. It should be highlighted that this characteristic 'change in the slope' is a common feature reported for the various systems infiltrated into porous media independently of the applied pore size and the materials the templates are made of (i.e., AAO or silica). [37,79] Moreover, it should be mentioned that the specific temperature (at which the change in the slope occurs) is reported to strongly depend on the pore size. In fact, the lower $d$, the higher the specific temperature is. This deviation of dynamics observed for infiltrated systems was recently assigned to two major effects: (1) changes in the dynamical heterogeneities, $\zeta$, within the glass formers under confinement conditions [68,80] and (2) the vitrification of the interfacial molecules. [79,81,82] The former approach implies that the pore size of applied porous membranes suppresses $\zeta$ upon cooling, resulting in the deviation of $\tau(T)$. However, this issue seems to be only valid for the systems characterized by $d\sim\zeta$, where $\zeta$ was calculated to approx. $\zeta\sim$2-3 nm.[68,80,83,84] Nevertheless, one can recall that the change of $\tau(T)$ was also reported for the cases where the applied porous templates were characterized by $d>$18 nm (significantly higher than $\zeta$). Therefore, this phenomenon is often described considering the vitrification of the interfacial layer (located near the interface). As a consequence that, below that temperature, the investigated system might be regarded as *pseudo*-isochoric, [85,86] which results in a change in the temperature dependence of relaxation times of core molecules, which follow $\tau_\alpha(T)$ at constant volume. Hence, the specific temperature (at which the change in the slope occurs) is usually assigned as the glass transition temperature of the interfacial molecules (located in close proximity to the interface and characterized by the 'higher' interfacial interactions and strongly reduced mobility), $T_{g,interfacial}$. Herein, we would like to point out three major observations. Firstly, the deviation of $\tau_{Dom}(T)$ can be observed for all applied types of membranes (native and modified). Although the change in the $\tau_{Dom}(T)$ occurs at different conditions (see **Table S1** in the **Supporting Information (SI) file**). Secondly, the specific temperature (at which the change in the slope of $\tau_{Dom}(T)$ occurs) agrees well with $T_{g,interfacial}$ determined form the calorimetric measurements, see **Figure 1(h)**. This implies that vitrification of the interfacial molecules might be responsible for the observed deviation of $\tau_{Dom}(T)$. Thirdly, it is worthwhile to stress that the temperature at which the $\tau_{Dom}(T)$ deviates from the bulk-like behavior does not correspond in any way to the behavior of the slower dielectric relaxation process appearing in the dielectric response of materials within native silica templates, see **Figure 3**. Note that the relaxation times of the additional dielectric process appearing in the case of infiltrated samples, which might be interpreted as the interfacial one, reaches $log_{10}\tau_{Add} \sim$ -2-(-3) when a change in the slope of the relaxation times of the main dielectric mode occurs. Taking into account also the results obtained from the calorimetric measurements, i.e., the presence of DGT and the fact that interfacial molecules form an interfacial layer, $\xi \sim 0.5$ nm, irrespectively of the character of pore walls, one can indeed assume that the deviation of $\tau_{Dom}(T)$ occurring at different conditions origins from the vitrification of the layer formed under applied confinement. However, the additional dielectric response observed in the loss spectra collected for PhAs within native silica templates of $d = 4$ nm (see **Figure 1**) is most likely not connected to the mobility of the interfacial layer.

Therefore, the question arises about the origin of the observed additional relaxation process. One can mention that the $\tau_{Add}(T)$-dependence resembles the characteristic of the slow

Arrhenius process (SAP) recently observed in the loss spectra collected for the (unequilibrated) polymeric thin films. [87] This molecular origin seems to be valid, especially taking into account that the activation energies, $E_a$, of this process determined for all examined samples reaches $E_{a,Add}$ = 84.5 kJ/mol (see **Table 2**), whereas those of SAP determined for a series of various polymeric thin films are of the order of $E_{a,SAP}$ = 100 kJ/mol. Nevertheless, it should be pointed out that present data are insufficient to clearly distinguish the origin of the additional dielectric mobility observed for the examined series of PhAs within native silica templates. Therefore, this issue would be explored further within the next projects.

**Table 2**. The activation energies, $E_{a,Add}$, calculated from the Arrhenius equation (Eq.4) for the additional (slower) relaxation mode observed in the dielectric response of PhAs incorporated into native silica membranes of $d$ = 4 nm.

|  | 2Ph1E | 3Ph1P | 4Ph1B | 5Ph1P |
|---|---|---|---|---|
| $E_{a,Add}$ [kJ/mol] | 84.5 | 83.9 | 84.5 | 69.9 |

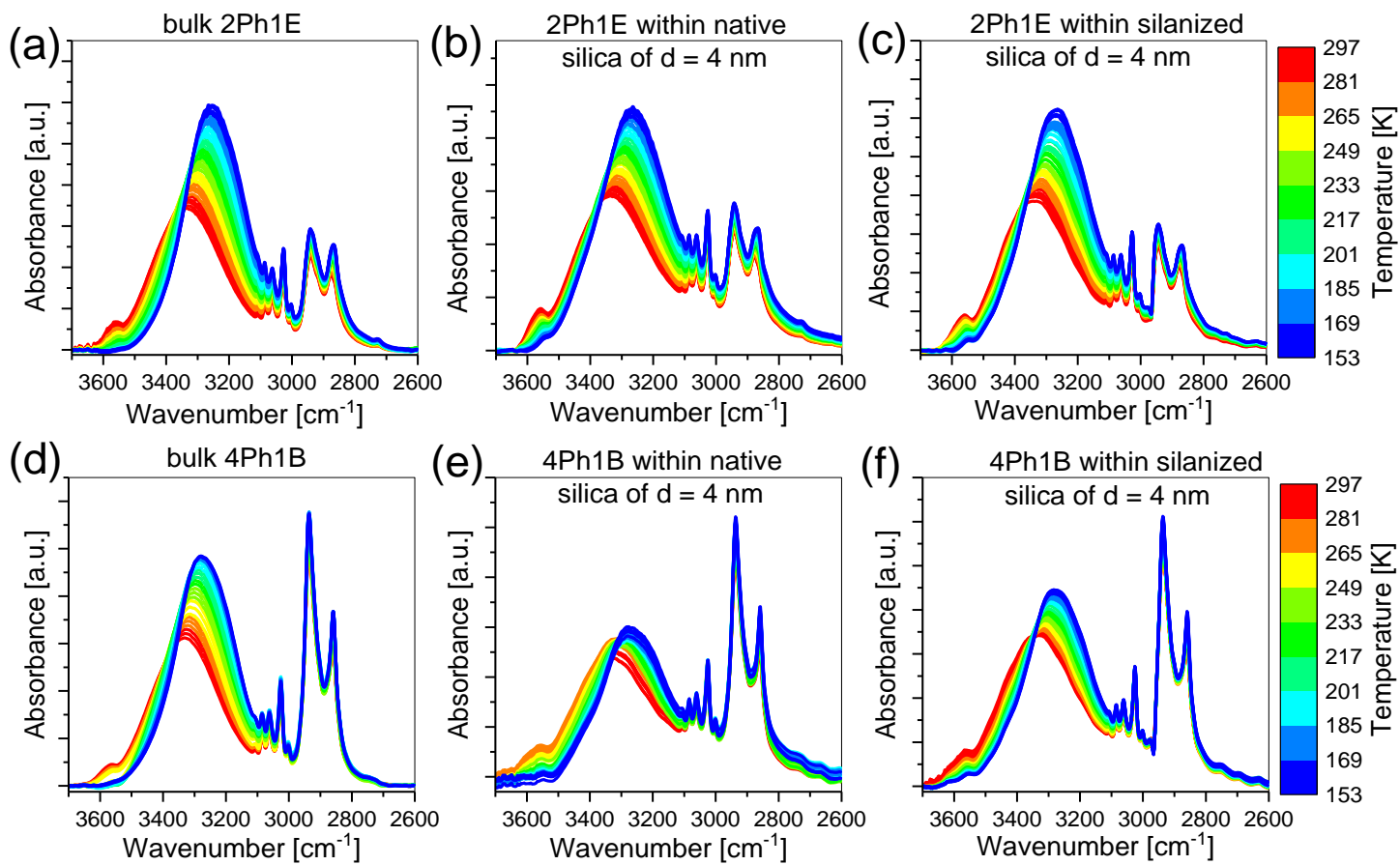

**Figure 4.** Infrared spectra of 2Ph1E in (a) bulk, (b) native pore, (c) silanized pore, and 4Ph1B in (d) bulk, (e) native pore, and (f) silanized pore in the frequency region 3700 – 2600 cm$^{-1}$ presented in the temperature range, $T$ = 298 – 153 K.

Alternatively, one can assume that this additional dielectric mode can originate from the mobility of 'new confinement-induced' associating structures. Note that in the case of the bulk materials, we do not observe any separate mobility related to the structural relaxation, but only the presence of a Debye-like process (related to the formation of nanoassociates). In fact, recent studies confirmed that the Debye-like process detected for PhAs is indeed related to the association-dissociation process according to the Transient Chain Model (TCM). [60] Thus, this additional mobility observed in pores might be related to the formation of 'new' confinement-induced associating structures. To shed new light on this problem, we performed infrared measurements to monitor the population of hydrogen bonds (HB) of the studied compounds under confinement. Herein, we focused mostly on monitoring the changes in H-bonding properties of PhAs, which can be well reflected/visible in the stretching vibration bands of hydroxyl groups located in the wavenumber range from 3700 to 2600 cm$^{-1}$. In this region, some peaks occur for both bulk and infiltrated PhA samples; namely, the band of weak intensity at ~ 3550 cm$^{-1}$ assigned to the stretching vibrations of the free, non-bonded O-H groups ($v_{OH\ free}$), and a broad signal centered at ~ 3330 cm$^{-1}$ associated with the stretching vibrations of H-bonded O-H moieties ($v_{OH\ bonded}$). At lower wavenumbers, the peaks related to the stretching vibrations of aromatic (3100 – 3000 cm$^{-1}$) and aliphatic (3000 – 2800 cm$^{-1}$) C-H groups are observed.

In **Figure 4**, IR spectra of bulk and infiltrated samples in the 3700 – 2600 cm$^{-1}$ spectral range over a wide temperature range ($T$ = 298 – 153 K) were shown. One should mention that the H-bonding properties of bulk PhAs were previously detailed in Ref. [44]; hence, in this paper, we mainly analyzed the associating behavior of infiltrated PhAs compared to the bulk ones. As shown in **Figure 4**, the most significant changes were observed in the $v_{OH}$ band region, as the position of this band was red-shifted (shifted to lower wavenumbers) with decreasing temperature in both bulk and confined PhAs. Such spectral behavior clearly indicates the strengthening of H-bonds in these systems upon cooling. The same effect was detected for other primary and secondary monohydroxy alcohols incorporated in silica and alumina mesopores [37] as well as water confined in periodic mesoporous (organo)silicas. [88] Furthermore, in both groups of systems, the intensity of the $v_{OH\ free}$ band decreased as the temperature was lowered/reduced, indicating the higher association degree (larger amount of H-bonded MA molecules).

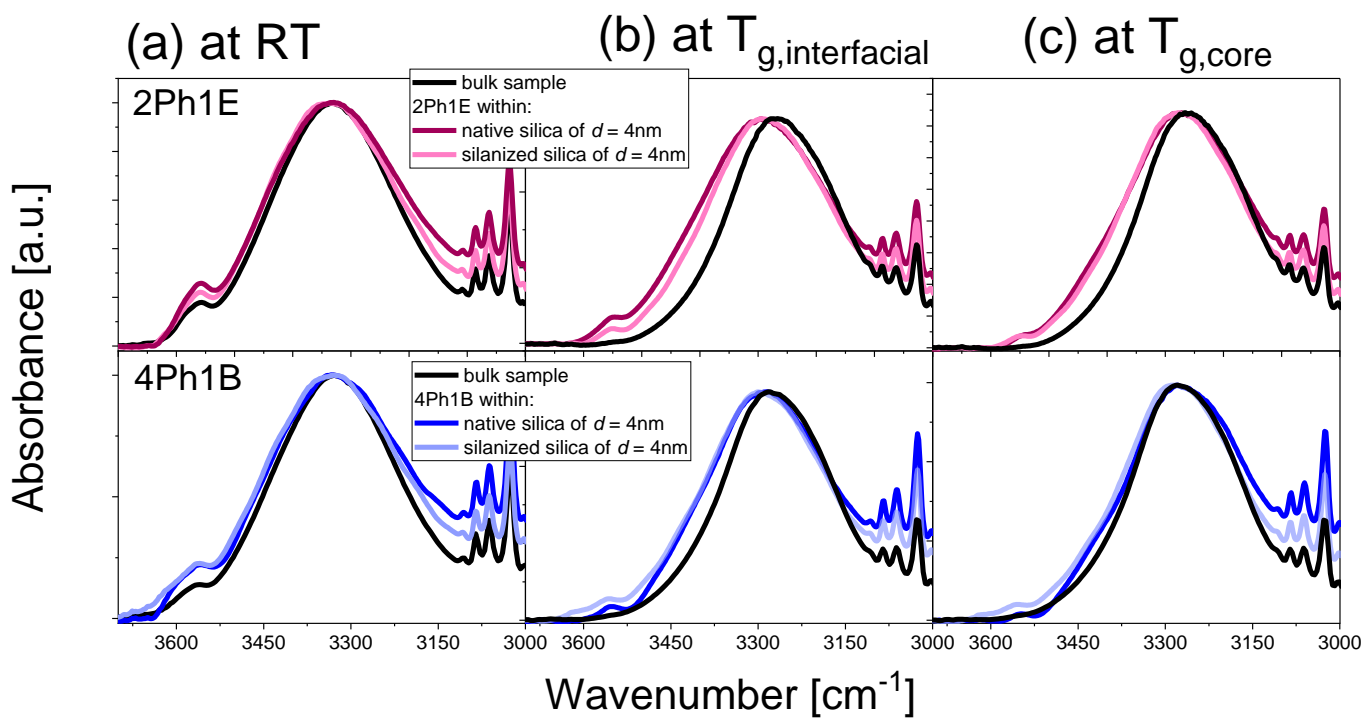

**Figure 5.** Comparison of FTIR spectra in the frequency region 3720 – 3000 cm$^{-1}$ for bulk and confined samples of examined PhAs at various chosen temperature conditions, (a) room temperature (RT), (b, a) calorimetric glass transition temperature of interfacial ($T_{g,interfacial}$), and core ($T_{g,core}$) molecules. For better clarity, the spectra were normalized to the OH stretching vibration ($\nu_{OH}$) band.

Furthermore, in **Figure 5**, we compared FTIR spectra of bulk and confined samples at three temperatures in the 3700 – 3000 cm$^{-1}$ spectral range. One can see that at RT, the $\nu_{OH}$ bands of bulk and confined PhAs exhibit nearly identical shapes, and their positions at maximum do not differ significantly (see **Table S2**). Interestingly, the spectroscopic studies conducted for water under confinement showed that the O-H stretching band was almost indistinguishable from bulk water [18,89], or it was only slightly blueshifted. [90] The reported blueshift phenomenon was associated with poorer H-bond acceptor ability of silica oxygen atoms and thus with the presence of interfacial water molecules. [90] Moreover, the intensity of the $\nu_{OH\ free}$ band was higher for confined PhA samples compared to the bulk. The percentage values of the number/amount of free hydroxyl groups for both bulk and spatially restricted PhAs were calculated and presented in **Figure S3**. It can be noticed that approximately twice as many non-associated OH groups for MAs under confinement compared to the non-confined samples is observed, indicating that the association process of PhAs in geometrical restriction is partly suppressed. These parameter values are similar for both native and silanized silica templates, with slightly higher values for silanized ones, suggesting that functionalizing silica membranes (more hydrophobic) and overall nanogeometrical restriction led to the presence of more amount of non-bonded OH species in nanopores. What is more, considerable changes in the $\nu_{OH}$ bandwidths are detected, i.e., the $\nu_{OH}$ bands of MAs in both native and silanized templates are broadened compared to their bulk counterparts. This indicates greater heterogeneity in the distribution of HB aggregates for alcohols under nanoconfinement. At RT, the broadening of this band is more distinct in the lower wavenumber region, suggesting that stronger H-bonds are more affected than weaker ones.

As the temperature decreased, the behavior of the $\nu_{OH}$ band resembles that at RT, i.e., the subtle structure of $\nu_{OH}$ band remains, with a weak peak originating from free hydroxyl groups, suggesting partial association at lower temperatures. As shown, the $\nu_{OH}$ peak for MAs in confined samples is shifted to higher wavenumbers, indicating weaker H-bonding interactions in spatial restriction (see **Figure 5(b)**). Moreover, significant discrepancies between confinement and bulk are observed in the $\nu_{OH}$ bandwidth, with confined samples having broader bands than bulk samples, especially at lower temperatures. The most prominent difference occurred on the left shoulder of the $\nu_{OH}$ band, which might indicate the presence of an additional contribution within the 'weaker-H-bonded' OH associate range, much more resolved when compared to the spectra at RT. Note that native silica nanopores exhibit slightly broader $\nu_{OH}$ bands compared to silanized membranes. Overall, the widening of the $\nu_{OH}$ band follows this order: bulk > silanized pore > native pore, implying a more substantial impact of the hydrophilic environment on the H-bond network of the incorporated PhAs. It should also be mentioned that similar observations were reported for other MAs under confinement, demonstrating that the incorporation of MAs into silica nanomaterials is manifested by the changes in the $\nu_{OH}$ peak frequencies, and the broadening of the band in the same order as described herein. [37]

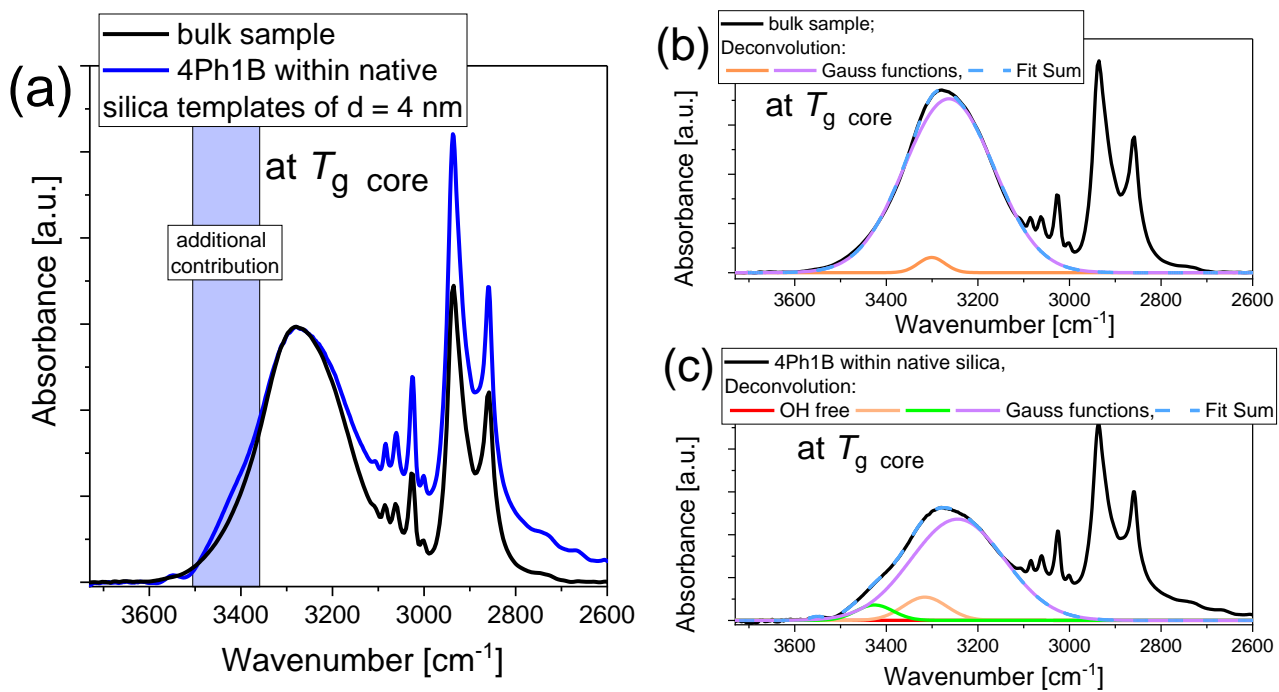

**Figure 6.** (a) The comparison of FTIR spectra of bulk 4Ph1B and the sample within native silica mesoporous templates at $T_{g,core}$. The spectra were normalized to the intensity of the OH band. The deconvolution of the $\nu_{OH}$ band of (b) bulk, and (c) infiltrated 4Ph1B into native silica pore at $T_{g,core}$ using Gauss functions.

To explore the presence of an additional shoulder in the higher frequency region observed for incorporated 4Ph1B (the blue box in **Figure 6(a)**), we carried out the deconvolution of these spectra recorded at $T_{g,core}$ in the range of 3600-3000 cm$^{-1}$. As found, the proper description of the $\nu_{OH}$ band of infiltrated 4Ph1B requires the application of three Gaussian curves (see **Figure 6(c)**), whereas the use of two Gaussian functions is enough to fit the $\nu_{OH}$ band of the bulk sample (**Figure 6(b)**). Interestingly, a new component occurring for the former sample in the higher wavenumber range suggests the existence of additional weak H-bond interactions occurring in the studied system. A similar deconvolution of the $\nu_{OH}$ band was presented for water confined in MCM-41 pores. In this case, the region ~ 3350 – 3500 cm$^{-1}$ was assigned to small aggregates of water molecules characterized by faster dynamics. [91] One can also add that mobility of this 'new' confined-induced HB interactions, resolved as an additional shoulder in IR spectra of incorporated PhAs, might contribute to the appearance of the additional relaxation process observed in loss spectra shown in **Figures 2(c,f)**.

**Table 3.** The calculated activation energies of dissociation, $E_a$, for PhAs incorporated into native and silanized silica membranes of $d = 4$ nm.

| Samples | $E_a$ [kJ/mol] | | |
|---|---|---|---|
| | Bulk PhAs[a] | PhAs within native silica of $d = 4$nm | PhAs within silanized silica of $d = 4$nm |
| 2Ph1E | 10.49 ± 0.28 | 6.94 ± 0.48 | 8.62 ± 1.39 |
| 3Ph1P | 11.64 ± 0.25 | 8.90 ± 0.52 | 7.96 ± 1.54 |
| 4Ph1B | 13.41 ± 0.80 | 8.84 ± 0.89 | 9.05 ± 0.74 |
| 5Ph1P | 13.65 ± 0.43 | 8.90 ± 0.88 | 8.47 ± 0.88 |

[a] Values of $E_a$ determined for the bulk samples were taken from Ref. [44].

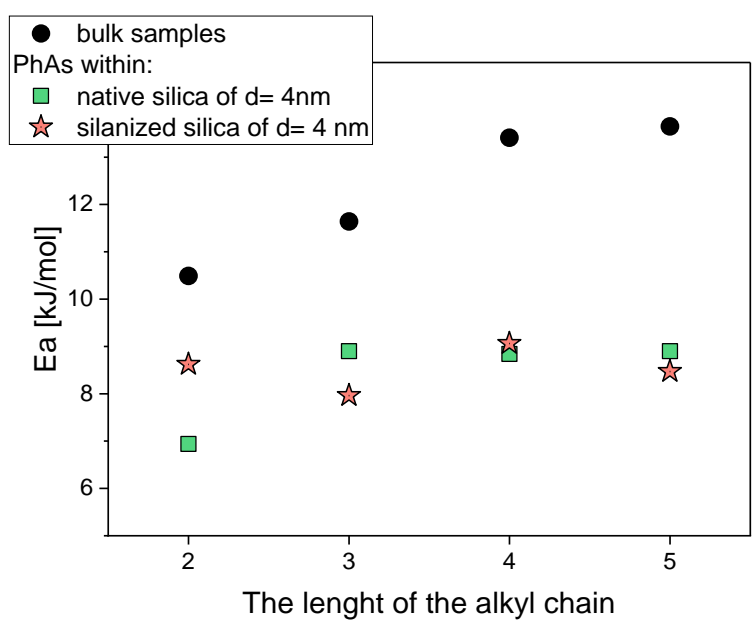

**Figure 7.** The activation energy of the dissociation, $E_a$, calculated for bulk PhAs and samples infiltrated into native and silanized silica templates of $d = 4$ nm.

Further, we performed the calculations of the activation energy, $E_a$, of the dissociation process, using the van't Hoff equation:

$$\ln K = -\frac{E_a}{RT} + \frac{\Delta S}{R}, \qquad (6)$$

where $E_a$ indicates the activation enthalpy, S is the entropy of the dissociation process, and R is the gas constant. The van't Hoff plots for PhAs in native mesopores are presented in **Figure S5**. As can be seen in **Figure 7**, the estimated values of $E_a$ for PhAs in nanorestriction are lower ($E_a$ ~7-9 kJ/mol) than in bulk materials (from 10.49 to 13.65 kJ/mol), indicating the significant contribution of weak H-bond interactions occurring in spatially confined geometries compared to the bulk ones in which the stronger HBs dominate. As a result, less energy is required to break these H-bonding interactions in confined samples than in bulk ones. This fact corresponds well with the lower association degree of infiltrated samples relative to that in bulk samples (slightly smaller number of H-bonded MA molecules relative to free ones under confinement compared to that in bulk samples). Moreover, as aliphatic chain length increases in bulk alcohols, dissociation energy tends to rise. However, this trend is not observed in confined samples, where $E_a$ values are comparable (refer to **Table 3** for values with errors). It should be mentioned that similar values of this parameter for bulk samples were reported for various aliphatic alcohols, differing in the chain length and the localization of the OH groups, with the activation energy of the dissociation process ranging from $E_a = 9 - 14$ kJ/mol. [92] Additionally, the literature shows that the enthalpy required to break HBs in pure water equals $E_a \sim 8$ kJ/mol, and a slightly higher value was calculated for HOD in the D$_2$O solution and is equal to $E_a \sim 10$ kJ/mol. [93] It should be stressed that, to the best of our knowledge, until now, the values of activation energy of the dissociation process for systems confined in nanopores have not been determined. Therefore, the values presented herein for PhAs in native and silanized nanomembranes make an important contribution to investigating the impact of nanorestrical confinement on the behavior of associating materials.

## IV. Conclusions

In this paper, we investigated the influence of surface interactions on the associating behavior of phenyl-substituted monohydroxyl alcohols. Interestingly, we observed a pronounced deviation of the dominant relaxation process (corresponding to the bulk Debye-like mode) for all examined confined systems occurring at $T_{g,interfacial}$, independently of the applied type of porous templates. Nevertheless, the dielectric response of PhAs infiltrated within native silica mesopores revealed the presence of an additional relaxation process, particularly pronounced for longer aliphatic chains. Interestingly, it was observed that this additional mobility is most likely not related to the vitrification of the interfacial layer, $T_{g,interfacial}$, but might have a similar origin as the SAP mode observed recently for various thin polymer films. Calorimetric data revealed a double glass transition phenomenon for systems infiltrated in native and modified silica. Moreover, it was found that for both types of membranes, the length of the interfacial layer, $\xi$, reaches approx. $\xi \sim 0.5$ nm for all PhAs. This suggests that an interfacial layer is formed irrespective of the character of the pore walls. This observation was explained considering contact angle measurements, which revealed that at low temperatures, all examined PhAs have a similar wettability on both surfaces, $\theta$ ~22-28º. IR measurements showed that the incorporation of PhAs into silica membranes inhibits a complete association over the entire temperature range. Moreover, nanogeometrical restriction has a relatively small impact on the H-bonds strength of infiltrated PhAs, as seen in the $\nu_{OH}$ peak position. However, it alters the $\nu_{OH}$ bandwidths, i.e., the confined samples are characterized by broader OH bands than those in bulk, indicating greater heterogeneity in the distribution of H-bonded systems in nanoconfinement. Notably, for the first time, we calculated the activation energy values of the dissociation process for confined PhAs, which were lower than those of bulk ones. This result correlates well with the lower association degree of infiltrated PhAs compared to their bulk counterparts, resulting from the spatial restriction. Thus, all experimental methods used consistently confirmed the formation of an additional interfacial layer in infiltrated PhAs in which the alcohol molecules strongly interact with the pore walls. We believe that the presented results offer a better understanding of the processes occurring for associating liquids in nanoconfinement.

**AUTHOR INFORMATION**

**Notes.** The authors declare no competing financial interests.

**Supporting Information**. The Supporting Information file is available free of charge at http://pubs.acs.org. It contains additional figures and tables, including figures of SEM pictures of native silica nanopore, the IR spectra of 'empty' silica templates, the contact angles measured for examined materials at three different temperatures, the percentages of free OH groups, the van't Hoff plots used to calculated activation energies of dissection processes, and tables containing glass transition temperatures and wavenumbers of the OH peaks.

**ACKNOWLEDGMENT.** NS, MT, and BH are thankful for financial support from the Polish National Science Centre (Dec. no 2019/33/B/ST3/00500).

**Author contributions:**

**N.S.** – writing – original draft, investigation, formal analysis, visualization

**M.T.** – writing – review & editing, conceptualization, investigation, funding acquisition

**B.H.** – writing – original draft, formal analysis, methodology

**M. G-R.** – investigation

**K.P.** – investigation

**K. K.** – supervision, writing – review & editing

https://doi.org/10.1002/cphc.200800616.

[24] K. Domin, K.Y. Chan, H. Yung, K.E. Gubbins, M. Jarek, A. Sterczynska, M. Sliwinska-Bartkowiak, Structure of ice in confinement: Water in mesoporous carbons, J. Chem. Eng. Data. 61 (2016) 4252–4260. https://doi.org/10.1021/acs.jced.6b00607.

[25] F. Corsetti, P. Matthews, E. Artacho, Structural and configurational properties of nanoconfined monolayer ice from first principles, Sci. Rep. 6 (2016) 1–12. https://doi.org/10.1038/srep18651.

[26] D.E. Moilanen, N.E. Levinger, M.D. Fayer, Confinement or properties of the interface? Dynamics of nanoscopic water in reverse micelles, Opt. InfoBase Conf. Pap. (2007) 8942–8948. https://doi.org/10.1364/ls.2007.ltuc3.

[27] A.J. Rieth, K.M. Hunter, M. Dincă, F. Paesani, Hydrogen bonding structure of confined water templated by a metal-organic framework with open metal sites, Nat. Commun. 10 (2019) 1–7. https://doi.org/10.1038/s41467-019-12751-z.

[28] T. Salez, Soft matter in confinement and at interfaces, Univ. Bordeaux. (2020).

[29] C. Weinberger, F. Zysk, M. Hartmann, N.K. Kaliannan, W. Keil, T.D. Kühne, M. Tiemann, The Structure of Water in Silica Mesopores – Influence of the Pore Wall Polarity, Adv. Mater. Interfaces. 9 (2022). https://doi.org/10.1002/admi.202200245.

[30] F. Bauer, R. Meyer, M. Bertmer, S. Naumov, M. Al-Naji, J. Wissel, M. Steinhart, D. Enke, Silanization of siliceous materials, part 3: Modification of surface energy and acid-base properties of silica nanoparticles determined by inverse gas chromatography (IGC), Colloids Surfaces A Physicochem. Eng. Asp. 618 (2021) 126472. https://doi.org/10.1016/j.colsurfa.2021.126472.

[31] A. Huwe, M. Arndt, F. Kremer, C. Haggenmüller, P. Behrens, Dielectric investigations of the molecular dynamics of propanediol in mesoporous silica materials, J. Chem. Phys. 107 (1997) 9699–9701. https://doi.org/10.1063/1.475265.

[32] R. Winkler, W. Tu, L. Laskowski, K. Adrjanowicz, Effect of Surface Chemistry on the Glass-Transition Dynamics of Poly(phenyl methyl siloxane) Confined in Alumina Nanopores, Langmuir. 36 (2020) 7553–7565. https://doi.org/10.1021/acs.langmuir.0c01194.

[33] A. Ghoufi, I. Hureau, D. Morineau, R. Renou, A. Szymczyk, Confinement of tert - Butanol Nanoclusters in Hydrophilic and Hydrophobic Silica Nanopores, J. Phys. Chem. C. 117 (2013) 15203–15212. https://doi.org/10.1021/jp404702j.

[34] M. Tarnacka, A. Talik, E. Kamińska, M. Geppert-Rybczyńska, K. Kaminski, M. Paluch, The Impact of Molecular Weight on the Behavior of Poly(propylene glycol) Derivatives Confined within Alumina Templates, Macromolecules. 52 (2019) 3516–3529. https://doi.org/10.1021/acs.macromol.9b00209.

[35] A. Górny, M. Tarnacka, S. Zimny, M. Geppert-Rybczyńska, A. Brzózka, G.D. Sulka, M. Paluch, K. Kamiński, Impact of Nanostructurization of the Pore Walls on the Dynamics of a Series of Phenyl Alcohols Incorporated within Nanoporous Aluminum Oxide Templates, J. Phys. Chem. C. 126 (2022) 18475–18489. https://doi.org/10.1021/acs.jpcc.2c05446.

[36] A. Ghoufi, I. Hureau, R. Lefort, D. Morineau, Hydrogen-bond-induced supermolecular assemblies in a nanoconfined tertiary alcohol, J. Phys. Chem. C. 115 (2011) 17761–

# Supporting material

# The impact of interface modification on the behavior of phenyl alcohols within silica templates

Natalia Soszka[a], Magdalena Tarnacka[b*], Barbara Hachuła[a],

Monika Geppert-Rybczyńska[a], Krystian Prusik[c], Kamil Kamiński[b]

[a] *Institute of Chemistry, University of Silesia in Katowice, Szkolna 9, 40-006 Katowice, Poland*

[b] *August Chełkowski Institute of Physics, University of Silesia in Katowice, 75 Pułku Piechoty 1a, 41- 500 Chorzów, Poland*

[c] *Institute of Materials Engineering, University of Silesia in Katowice, 75 Pułku Piechoty 1a, 41-500 Chorzów, Poland*

*Corresponding author: magdalena.tarnacka@us.edu.pl

## Table of content

**Scanning Electron Microscope (SEM).** The pores distribution images were taken by JEOL JSM-7100F TTLs LV/EDS field-emission scanning electron microscope (SEM) operated at 10 kV acceleration voltage.

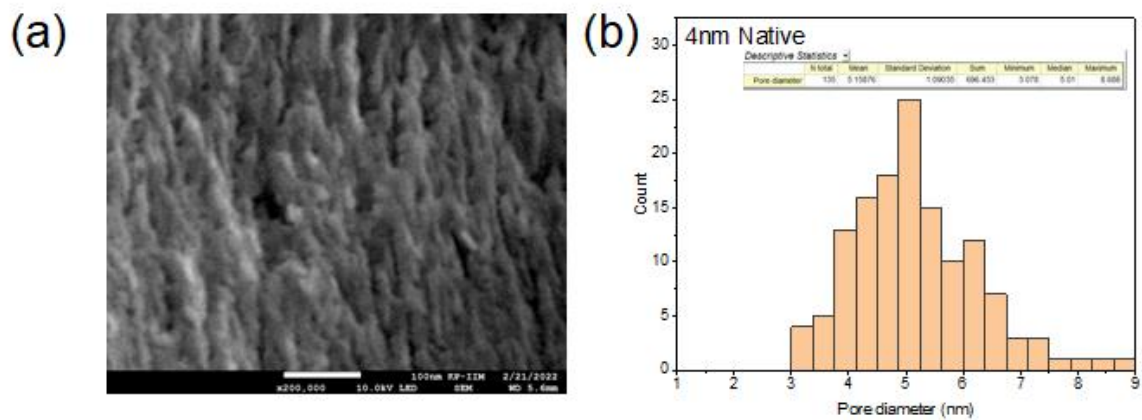

**Figure S1.** SEM image of native silica nanopore ($d$ = 4 nm) in (a) cross section view and (b) histogram with the estimation of the diameter of the pore.

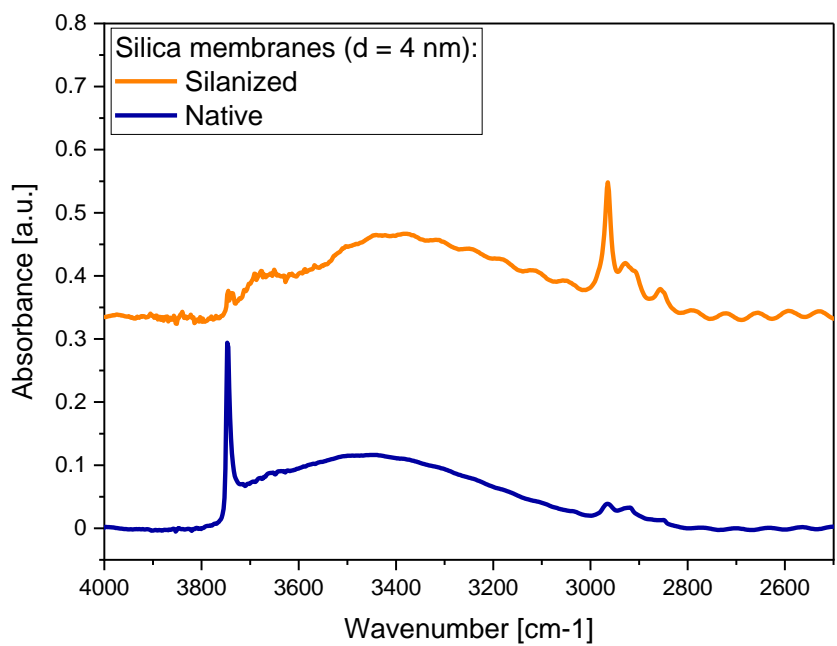

**Figure S2.** Infrared spectra of „empty" native (blue) and silanized (orange) silica templates (d = 4 nm) in the frequency range 4000 – 2500 cm$^{-1}$.

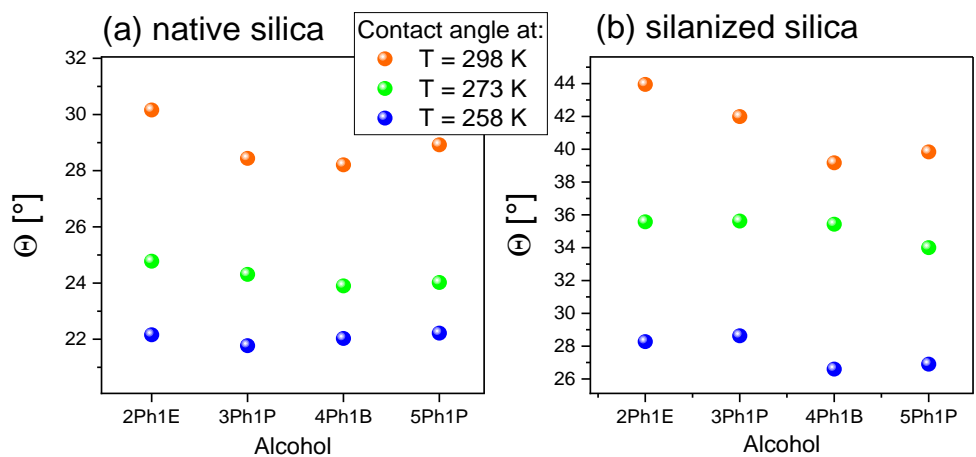

**Figure S3.** The alkyl chain dependences of contact angle, θ, measured for all examined materials at three different temperatures.

**Table S1.** Comparison of the glass transition temperatures obtained from BDS and DSC measurements, for all studied samples (both bulk and confined ones). The uncertainty of $T_g$ determination is ± 2 K for all presented values. Data for PhAs within AAO of $d = 10$ nm were taken from Ref. [1].

| Sample | DSC measurements[a] | | BDS measurements[b] | |
|---|---|---|---|---|
| | $T_{g,core}$ [K] | $T_{g,interfacial}$ [K] | $T_{g,interfacial}$ [K] | $T_{g,core}$ [K] |
| **2Ph1E** | | | | |
| bulk | - | 184.4 | - | 180.4 |
| Native silica of $d = 4$ nm | 171.7 | 217.9 | 205 | 166 |
| Silanized silica of $d = 4$ nm | 174.1 | 206.6 | | |
| AAO of $d = 10$ nm | 180.4 | 210.2 | 197.2 | 162.0 |
| **3Ph1P** | | | | |
| bulk | - | 181.9 | - | 178.4 |
| Native silica of $d = 4$ nm | 166.2 | 214.8 | 209 | 165 |
| Silanized silica of $d = 4$ nm | 171.3 | 205.5 | | |
| AAO of $d = 10$ nm | 176.1 | 208.4 | 201.2 | 165.2 |
| **4Ph1B** | | | | |
| bulk | - | 180.9 | - | 175.2 |
| Native silica of $d = 4$ nm | 165.5 | 212.9 | 207 | 161 |
| Silanized silica of $d = 4$ nm | 169.3 | 205.5 | | |
| AAO of $d = 10$ nm | 175.5 | 211.4 | 199.2 | 166.6 |
| **5Ph1P** | | | | |
| bulk | - | 181.8 | - | 177.8 |
| Native silica of $d = 4$ nm | 167.1 | 215.3 | 210 | 159 |
| Silanized silica of $d = 4$ nm | 171.8 | 204.9 | | |
| AAO of $d = 10$ nm | 176.2 | 208.2 | 199.2 | 165.4 |

[a] Note that calorimetric $T_g$s were determined from the midpoint of the observed heat capacity jumps;
[b] Dielectric $T_g$s were estimated according tot the procedure listed in **Material and Methods**.

**Table S2.** Wavenumbers of the OH peak for measured PhAs (bulk and infiltrated into nanopores) at room temperature (RT), and at higher and lower glass transition temperatures.

| Alcohol | Wavenumber [cm$^{-1}$] at RT | Wavenumber [cm$^{-1}$] at $T_{g\ interfacial}$ | Wavenumber [cm$^{-1}$] at $T_{g\ core}$ |
|---|---|---|---|
| bulk | | | |
| 2Ph1E | 3330 | 3285 | 3262 |
| 3Ph1P | 3329 | 3285 | 3268 |
| 4Ph1B | 3328 | 3287 | 3274 |
| 5Ph1P | 3330 | 3293 | 3280 |
| native (unmodified) SiO$_2$ templates of d= 4 nm | | | |
| 2Ph1E | 3329 | 3291 | 3273 |
| 3Ph1P | 3328 | 3291 | 3279 |
| 4Ph1B | 3333 | 3286 | 3272 |
| 5Ph1P | 3327 | 3298 | 3274 |
| silanized (modified) SiO$_2$ templates of d = 4 nm | | | |
| 2Ph1E | 3337 | 3291 | 3277 |
| 3Ph1P | 3329 | 3288 | 3263 |
| 4Ph1B | 3328 | 3297 | 3285 |
| 5Ph1P | 3326 | 3294 | 3282 |

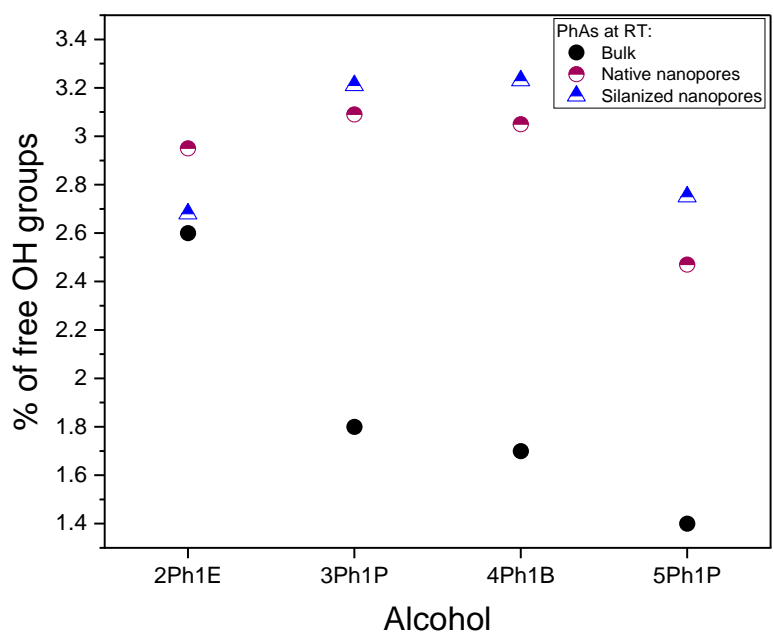

**Figure S4.** The percentage of free OH groups estimated for bulk PhAs (black), and infiltrated into native (burgundy) and silanized (blue) silica nanopores ($d$ = 4 nm).

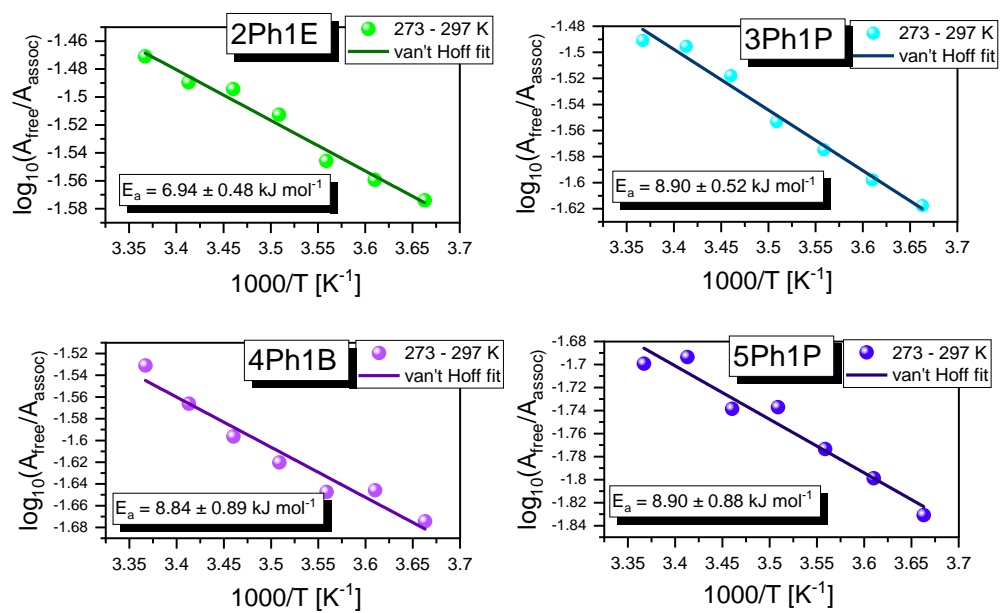

**Figure S5.** The van't Hoff plots for measured PhAs in native silica mesopores (d = 4 nm) used to calculate the activation energy of dissociation process.